\documentclass[twocolumn,aps,prl,showpacs,superscriptaddress]{revtex4}
\usepackage{color}
\usepackage{mathtools}
\usepackage{graphicx}
\usepackage{SIunits}

\begin{document}

\title{{Dynamical Gate Tunable Supercurrents in Topological Josephson Junctions}}

\author{C. Kurter}
\affiliation{Department of Physics, University of Illinois at Urbana-Champaign, Urbana, IL 61801}
\affiliation{Frederick Seitz Materials Research Laboratory, University of Illinois at Urbana-Champaign, Urbana, IL 61801}

\author{A. D. K. Finck}
\affiliation{Department of Physics, University of Illinois at Urbana-Champaign, Urbana, IL 61801}
\affiliation{Frederick Seitz Materials Research Laboratory, University of Illinois at Urbana-Champaign, Urbana, IL 61801}

\author{P. Ghaemi}
\affiliation{Department of Physics, University of Illinois at Urbana-Champaign, Urbana, IL 61801}
\affiliation{Physics Department, City College of the City University of New York, New York, NY 10031}

\author{Y. S. Hor}
\affiliation{Department of Physics, Missouri University of Science and Technology, Rolla, MO 65409}

\author{D. J. Van Harlingen}
\affiliation{Department of Physics, University of Illinois at Urbana-Champaign, Urbana, IL 61801}
\affiliation{Frederick Seitz Materials Research Laboratory, University of Illinois at Urbana-Champaign, Urbana, IL 61801}

\date{\today}

\begin{abstract}

Josephson junctions made of closely-spaced conventional superconductors on the surface of 3D topological insulators have been proposed to host Andreev bound states (ABSs) which can include Majorana fermions. Here, we present an extensive study of the supercurrent carried by low energy ABSs in Nb/Bi$_2$Se$_3$/Nb Josephson junctions in various SQUIDs as we modulate the carrier density in the Bi$_2$Se$_3$ barriers through electrostatic top gates. As previously reported, we find a precipitous drop in the Josephson current at a critical value of the voltage applied to the top gate. This drop has been attributed to a transition where the topologically trivial 2DEG at the surface is nearly depleted, causing a shift in the spatial location and change in nature of the helical surface states. We present measurements that support this picture by revealing qualitative changes in the temperature and magnetic field dependence of the critical current across this transition. In particular, we observe pronounced fluctuations in the critical current near total depletion of the 2DEG that demonstrate the dynamical nature of the supercurrent transport through topological low energy ABSs.  
\end{abstract}


\maketitle

Topological insulators have robust conducting surface states protected by time reversal-symmetry~\cite{RevModPhys.82.3045}. When conventional superconducting leads form a lateral Josephson junction on the surface of a topological insulator, the constructive interference of electron and hole-like excitations forms Andreev bound states (ABSs) whose energy spectrum is sensitive to the relative phase difference $\phi$ between the superconducting leads. Proximity-induced supercurrent flowing through the topological insulator segment of the junction is carried by these ABSs. Such junctions are proposed to be a platform to realize and manipulate Majorana bound states~\cite{Majorana1937} which are zero energy ABSs at phase bias $\pi$. The Majorana modes can store and process quantum information nonlocally through their non-Abelian exchange statistics; therefore, they are a key component in a fault-tolerant topological quantum computer~\cite{RevModPhys.80.1083}. 

Motivated by the theoretical proposals in Ref~\cite{PhysRevLett.100.096407}, several groups have experimentally demonstrated proximity-induced superconductivity in 3D topological insulators such as  Bi$_2$Se$_3$ and Bi$_2$Te$_3$~\cite{PhysRevB.84.165120, NatCommun.2.575, NatMat.11.417, PhysRevLett.109.056803, SciRep.2.339, PhysRevB.85.045415, NatCommun.4.1689, PhysRevX.3.021007, Kurter2013, Orlyanchik2013, PhysRevB.89.134512}. By changing the chemical potential in the topological insulator with an electrostatic gate, the supercurrent can be tuned \cite{NatCommun.2.575, NatCommun.4.1689, Kurter2013, Orlyanchik2013}, suggesting that the supercurrent flows primarily through surface states rather than through the bulk \cite{NatMat.11.417}. Although the conducting bulk of common topological insulators might bring serious doubts in this regard, it was theoretically shown that the Andreev states in the bulk of topological insulators have higher energy than the Andreev states formed out of the surface states, except at a critical doping level at which a dramatic transition changes the location and nature of the surface states~\cite{PhysRevLett.107.097001,PhysRevLett.109.237009,PhysRevB.87.035401}. Recently, a sharp drop in the supercurrent of a Nb/Bi$_2$Se$_3$/Nb junction at a specific value of gate voltage was reported and explained in terms of this phenomenon \cite{Orlyanchik2013}.

In this paper, we present an experimental study on devices made of Nb/Bi$_2$Se$_3$/Nb junctions which exhibit the presence of a regime where supercurrent predominantly flows through the helical surface states in the Bi$_2$Se$_3$. The data suggests complex changes in the distribution of the supercurrent as the junctions are gated that are important in understanding the transport and phase-sensitive Josephson properties of hybrid superconductor-topological insulator structures. 

Our devices incorporate lateral Josephson junctions fabricated on top of the 3D topological insulator Bi$_2$Se$_3$.  We exfoliate Ca-doped Bi$_2$Se$_3$ (Bi$_{1.9975}$Ca$_{0.0025}$Se$_{3}$) flakes onto doped Si substrates with a 300 nm thick oxide layer. Intercalation of Ca atoms reduces the high n-doping in as-grown Bi$_2$Se$_3$ so that the bulk transport is suppressed, but likely not totally eliminated~\cite{Nature.460.1101, PhysRevLett.106.196801}.  After thin, large area, and flat pieces are identified with atomic force microscopy, we define the junction leads by e-beam lithography and lift-off of a dc sputtered film of 60 nm Nb following \emph{in situ} Ar ion-milling of the sample surface.

\begin{figure}
\centering
\includegraphics[bb=14 190 582 750,width= 3 in]{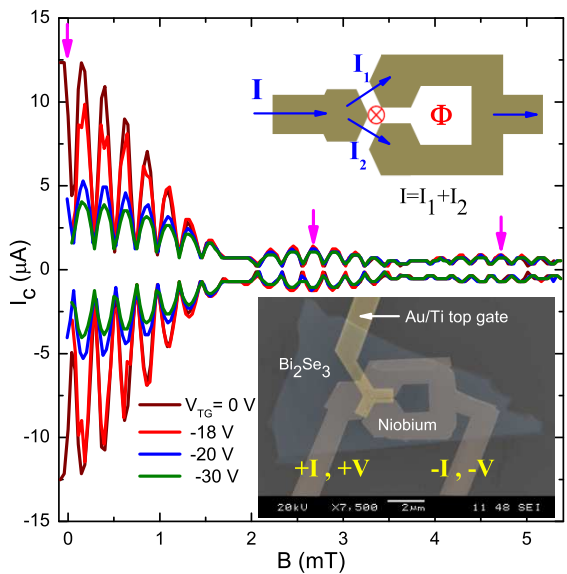}
\caption{(Color online) Main: The evolution of the Fraunhofer-like diffraction envelope enclosing SQUID oscillations with top gate bias. Top inset: Schematic of the tri-junction SQUID, three superconducting leads meeting at a tri-junction on the topological insulator; gates are not shown in the picture. The total supercurrent I divides into I$_1$ and I$_2$, the supercurrents passing through junction 1 and junction 2 respectively. The third junction is not directly probed here. When an out-of-plane magnetic field is applied, the flux $\Phi$ threading the loop gives rise to rapid SQUID oscillations bounded by the diffraction patterns of the individual junctions. Bottom inset: False-color scanning electron microscope image of the tri-junction SQUID fabricated on a 86 nm thick Bi$_2$Se$_3$ flake.} \label{fig:overview}
\end{figure}

The schematic and an SEM image of one such device are shown in the insets of Fig.~\ref{fig:overview}. This uses a special tri-junction SQUID configuration in which three closely-placed Nb leads (separated by approximately 100 nm) meet at a tri-junction on the surface of an 86 nm-thick exfoliated flake of a Bi$_2$Se$_3$ crystal. Two of the superconducting leads are connected in a loop so that their relative phase difference can be tuned by applying an out-of-plane magnetic field. We measure the transport between a contact on the loop and the third superconducting lead. An electrostatic top gate (not shown in the schematic but SEM image) is fabricated by covering the sample with a ~33 nm Al$_2$O$_3$ layer via atomic layer deposition and then depositing a Ti/Au electrode. By applying negative bias to the electrostatic top gates, we can study the junctions in various regimes. The samples are mounted on a copper cold finger that is bolted to the mixing chamber of a dilution refrigerator.  Unless otherwise stated, all measurements were performed at 10 mK.

Figure~\ref{fig:overview} shows the out-of-plane magnetic field modulation of the supercurrent at various gate voltages. It exhibits dc-SQUID oscillations arising from the magnetic flux threading the loop, bounded by an envelope set by the interference pattern of the individual Josephson junctions. The third junction only modulates the circulating current in the SQUID loop and is not directly probed here.  Because the current flowing across the third junction is at least two orders of magnitude smaller than the current circulating around the SQUID loop (i.e. $I_{c3} \ll \Phi_0 / L$, where $I_{c3}$ is the critical current of the third junction, $L$ is the SQUID loop inductance, and $\Phi_0 = h / 2e$ is the magnetic flux quantum), the third junction only weakly perturbs the phase difference between the two probed junctions and can be ignored here. There is a slight offset in the location of the maximum critical current at zero flux which we attribute to the residual field from our magnet. 

\begin{figure}
\centering
\includegraphics[bb=9 117 590 708,width= 3.5 in]{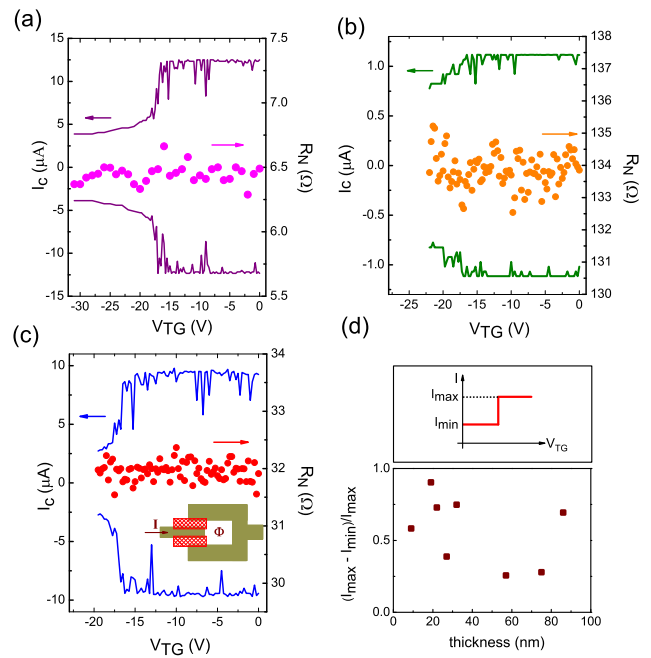}
\caption{(Color online) Critical current I$_c$ (solid curve) and normal state resistance R$_N$ (dots) vs.~top gate bias V$_{TG}$ for (a) the tri-junction SQUID shown in Fig.~1 (b) another tri-junction SQUID fabricated onto a 57 nm-thick Bi$_2$Se$_3$ flake and (c) a dc-SQUID on a 22 nm Bi$_2$Se$_3$ flake; the inset shows the schematic of the dc-SQUID incorporating two top gated Nb/Bi$_2$Se$_3$/Nb junctions through a superconducting loop. (d) The fractional change $(I_{max} - I_{min})/I_{max}$ of the critical current between the intrinsic density (I$_{max}$) and after the 2DEG is completely depleted (I$_{min}$) as a function of crystal thickness for different devices based on Nb/Bi$_2$Se$_3$/Nb junctions. The sketch in the top panel of (d) shows how I$_{max}$ and I$_{min}$ are determined. } \label{fig:topgate}
\end{figure}

Fig.~\ref{fig:topgate}a shows the effect of the top gate bias V$_{TG}$ on the critical current I$_c$ and normal state resistance $R_N$ of the tri-junction device shown in Fig.~\ref{fig:overview}. For the Bi$_2$Se$_3$ crystals we use here, ARPES \cite{Nature.460.1101} and transport measurements \cite{PhysRevLett.106.196801} show that the chemical potential resides in the conduction band at the intrinsic density. When we apply negative bias to the top gates, the critical current is almost independent of V$_{TG}$ over a wide range. We then observe a sharp drop in critical current beyond a particular V$_{TG}$ of $\sim$ -18 V, giving way to a gentle decline. Over the same gating range, the normal state resistance is virtually constant suggesting that the conductance is dominated by bulk transport whereas the majority of the supercurrent transport is through the surface states \cite{NatCommun.4.1689}.

We have observed similar behavior in another tri-junction SQUID fabricated on a 57 nm-thick Bi$_2$Se$_3$ crystal (see Fig.~\ref{fig:topgate}b) as well as a dc-SQUID made of 22 nm-thick crystal (see Fig.~\ref{fig:topgate}c). Single Josephson junctions built either on Bi$_2$Se$_3$ films~\cite{Orlyanchik2013} or exfoliated crystals~\cite{Kurter2013} also demonstrated similar characteristics, so we believe the behavior to be intrinsic to the junctions and independent of the device geometry.

This sharp drop observed in I$_c$ vs. V$_{TG}$ data has been proposed to arise from a  transition occurring at a critical value of the chemical potential, just prior to the depletion of the conventional 2DEG~\cite{Orlyanchik2013}. At the transition, the ABSs in the 2DEG become gapless and the helical ABSs physically move from the bottom of the band-bended region to the top surface. Although we have no direct evidence for the shift in the location of the supercurrent-carrying states, it is expected that it will result in a reduction in critical current due to a change in the transport properties of the helical surface states, which we do observe. It is important to note that in addition to the band bended regions and topological surface states, there are conducting channels in the bulk of the topological insulators. As a result, it is possible that part of the supercurrent is carried through the bulk of the sample. But as the gate is screened before affecting the bulk states, we do not believe that they play much role in the sudden drop in critical current with gating. Moreover, the poor connectivity of bulk states as well as the higher energy of their Andreev states as predicted theoretically \cite{PhysRevLett.107.097001,PhysRevLett.109.237009,PhysRevB.87.035401} point toward their low contribution in the supercurrent in the junction.

One can demonstrate the overall unimportance of the bulk to the Josephson current by showing that the gate-tuned transition that we observe is independent of sample thickness.  Although the critical current changes from device to device due to varitions in junction length, junction width, and sample quality, it is instructive to consider the thickness dependence of the fractional change in the critical current as each device is gate-tuned from its intrinsic carrier density (denoted as $I_{max}$) to a regime where 2DEG is completely depleted (denoted as $I_{min}$).  In Fig.~\ref{fig:topgate}d, we plot the fractional change in the supercurrent ($I_{max}-I_{min})/ I_{max}$ vs. Bi$_2$Se$_3$ thickness for all of the gated samples that we studied. We find that this quantity is at most only weakly dependent on thickness. Since we expect that the top gate primarily tunes the carrier density of the surface states due to screening from bulk states near the surface, any supercurrent carried by the bulk should be roughly gate-independent but increases with sample thickness.  Likewise, any supercurrent carried by the surface states is likely to be strongly gate-dependent but only weakly dependent on sample thickness.  Thus, the sharp drop we observe with gating should reflect a change in the supercurrent carried by the surface states, with at most a gate-independent background from the bulk states. This supports the hypothesis that the supercurrent predominantly carried by the surface states rather than the bulk, as asserted by earlier works \cite{NatMat.11.417, NatCommun.4.1689, PhysRevB.89.134512}.  Thus, the primary effect of the top gate is to tune the supercurrent from surface states.

\begin{figure}
\centering
\includegraphics[bb=1 57 589 663,width= 3 in]{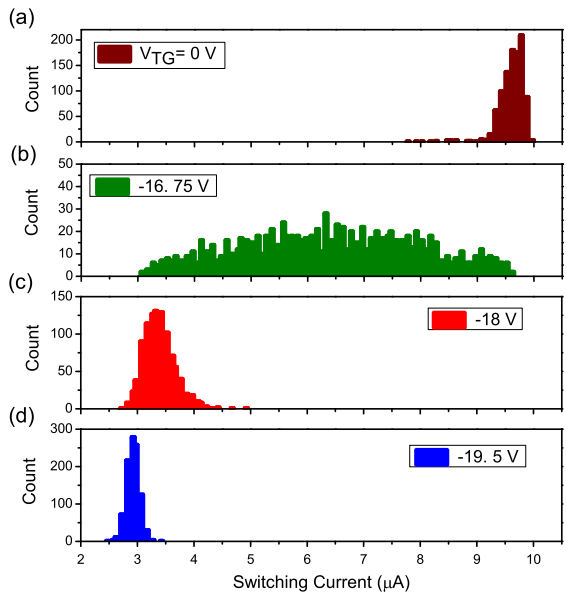}
\caption{(Color online) (a)-(d) Switching current distribution for the 22 nm-thick dc-SQUID including two Nb/Bi$_2$Se$_3$/Nb junctions at various top gate biases. Note that the distribution is anomalously broad at V$_{TG}$= -16.75 V, which is the onset of the transition; see Fig.~\ref{fig:topgate}c.} \label{fig:switching}
\end{figure}

A striking feature of the gate dependence is that as we increase the negative gate bias and approach the transition, the critical current shows pronounced downward fluctuations; these largely disappear once the transition is passed.  To study this, we measured the current at which the device switches into the finite voltage state in successive ramps of the bias current.  We perform 1000 ramps at a rate of $\sim$ 0.1 Hz.  In Fig.~\ref{fig:switching}, we show the distributions of these switching currents in the 22 nm-thick dc-SQUID shown in Fig.~\ref{fig:topgate}c for four top-gate values. Far above and below the transition, the distributions are narrow with tails that indicate fluctuation-induced switches that become more frequent as the transition is approached. Near the transition, the distribution is very wide, essentially encompassing all possible values of switching current between the two extreme values of critical current.

The switching distributions suggest that different regions of the junction switch at different values of the bias current.  This inhomogeneous switching is mostly likely a result of charge noise, which leads to spatially and temporally random electric fields that can cause certain areas of the junction to switch from one state to the other. The origin of this charge noise could either be fluctuators within the Al$_2$O$_3$ dielectric \cite{RevModPhys.53.497, RevModPhys.60.537} or charged defects in the Bi$_2$Se$_3$ crystals.  Although we have not tried to make a comprehensive dynamical model for the critical current, we would postulate that the local transitions are thermally-activated and depend on the local gate voltage that is subject to the charge distribution. It is also likely that there are long-range interactions between regions induced by the extended ABSs; that is suggested by the number of large suppressions in the critical current observed for gate voltages far above the gate voltage at which the time-averaged critical current drops, apparent in Fig.~\ref{fig:topgate}a-c. The resulting time-variation in critical current as a function of gate voltage, temperature, and magnetic field ultimately depends on considering the local charge fluctuations and percolation physics.


\begin{figure}
\centering
\includegraphics[bb=6 105 580 626,width= 3 in]{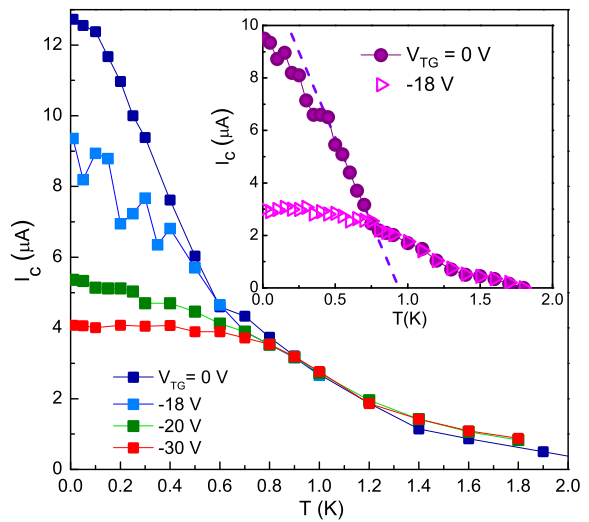}
\caption{(Color online) Temperature dependence of the critical current in the 86 nm-thick tri-junction device at various top gate bias and at zero field.  Inset: Temperature dependence of critical current in the 57 nm-thick dc-SQUID at zero field.  Dashed line shows a change in slope associated with kink.} \label{fig:temperature}
\end{figure}

We further explore the gate-tuned transition by considering the effects of temperature and magnetic field on the supercurrent. The main panel of Fig.~\ref{fig:temperature} shows critical current I$_c$ vs.~temperature $T$ for various values of top gate bias V$_{TG}$. At zero gate bias, the critical current drops rapidly with increasing temperature with an overall upward curvature. Such behavior had been previously interpreted as a signature of ballistic supercurrents in topological insulator Josephson junctions~\cite{NatMat.11.417}. However, in our devices there is an abrupt change of slope in the V$_{TG}$ = 0 trace at $\sim$ 600 mK, suggesting a more complicated origin.  As top gate bias is decreased, we deplete the 2DEG states in the Bi$_2$Se$_3$ and the critical current drops, saturating for V$_{TG}> -30$ V. At this extreme, critical current is almost unchanged for temperatures between 0 and 750 mK, then merges with the zero voltage curve at the temperature of the kink feature. This temperature dependence is consistent with the behavior of a diffusive SNS junction \cite{NatMat.11.417}.  We see this behavior in almost all of our devices, such as in the inset of Fig.~\ref{fig:temperature} for the 22 nm-thick dc-SQUID in which the kink and merging of the curves are clearly seen. In the intermediate regime at gate voltages near V$_{TG}$ = -18 V, where the sudden drop of the critical current occurs, the critical current exhibits large variations with temperature as a result of the broad distribution in the critical switching current. For this gate voltage, as the temperature increases the width of the switching distribution narrows and the critical current merges with the zero gate bias curve.

This temperature behavior is consistent with the model used to explain the drop in the supercurrent in Ref~\cite{Orlyanchik2013}. In this model, both the topologically trivial 2DEG and non-trivial helical surface states can carry supercurrent. The key feature of the Andreev states formed in the 2DEG is that their energy depends on the position of the chemical potential\cite{PhysRevLett.107.097001,PhysRevLett.109.237009,PhysRevB.87.035401}. The energy of these 2DEG ABSs is in general larger than the energy of the ABSs formed out of the helical surface states~\cite{Orlyanchik2013}. Thus, the 2DEG states at low temperature contribute only a minor amount of supercurrent, as evident from the insensitivity of critical current to changes in top gate bias prior to the critical gate voltage. Instead, the 2DEG primarily serve to screen out disorder that might be deleterious to the topologically non-trivial surface states. At the transition, the energy of 2DEG ABSs becomes equal to the helical state ABSs and they provide the channel for the displacement of the helical ABSs to the surface. This changes the nature of topological insulator barrier in the junction by altering the transmission coefficient. The difference in the transport properties of the layer which accommodates the low energy ABSs is apparent in the temperature dependence of the supercurrent before and after the transition at low temperatures. In this model, the kink is then associated with the presence of two sets of ABSs, one formed out of the band-bended region and one formed out of the helical surface states. Notice that these states can have different induced gap size and critical temperature as well as different transmission probability at the lids. 

\begin{figure}
\centering
\includegraphics[bb=100 240 446 661,width= 2.6 in]{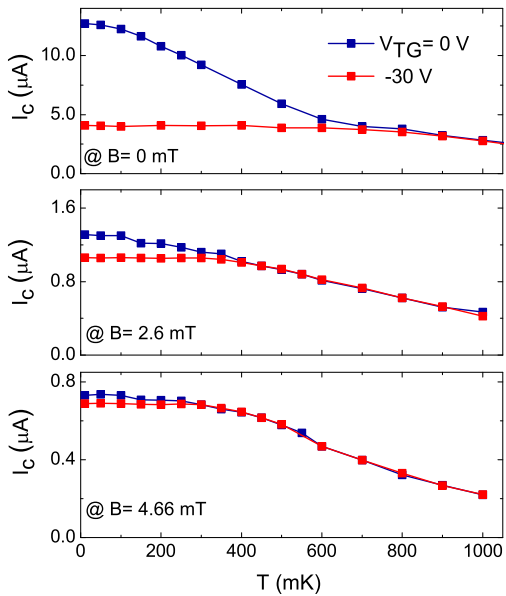}
\caption{(Color online) Temperature dependence of critical current at various magnetic fields for the 86 nm-thick tri-junction device before and after the transition. Note that the temperature at which the curves merge decreases with increasing magnetic field.} \label{fig:temperature2}
\end{figure}

In Fig.~\ref{fig:temperature2} we examine how the temperature dependence of the supercurrent evolves with magnetic field.  The field values correspond to the maximum critical current of the lobes of magnetic diffraction envelope (shown with pink arrows in Fig.~\ref{fig:overview}). For each field we plot two extreme regimes: at zero gate bias, where supercurrent is carried by multiple channels, and at V$_{TG}$= -30 V, where the 2DEG is completely depleted. We find that at higher field the $V_{TG}=0$ curve mimics the diffusive behavior of the V$_{TG}$= -30 V.  Note that the converging temperature shrinks with magnetic field, suggesting that the transition at which helical states moved to the top of the surface is washed out by the magnetic field.

\begin{figure}
\centering
\includegraphics[bb=33 225 558 716,width= 3 in]{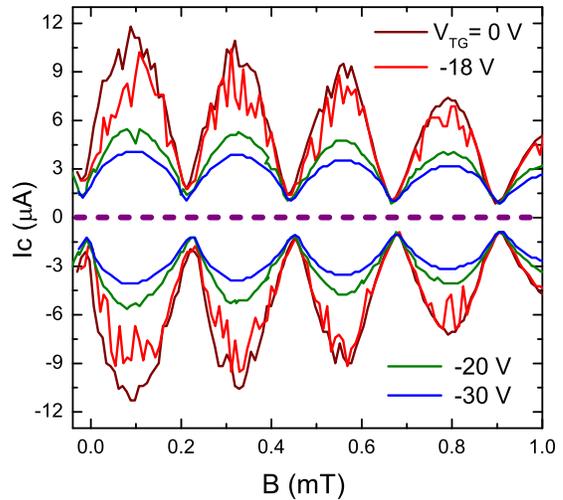}
\caption{(Color online) SQUID oscillations of the 86 nm thick tri-junction device for four different top gate biases. The critical top gate bias V$_{TG}$= -18 V demonstrates that the appearance of fluctuations in the supercurrent are consistent with the previous data sets.} \label{fig:SQUIDosc}
\end{figure}

We can also see the erratic fluctuations as a precursor for aforementioned transition in magnetic field dependence of the supercurrent. Figure~\ref{fig:SQUIDosc} shows the rapid oscillations of the critical current with a periodicity of one flux quantum threading the SQUID loop, bounded by the Fraunhofer-like single junction critical current modulation (see Fig.~\ref{fig:overview} for the diffraction envelope showing 3 lobes) for four  V$_{TG}$ values. At the critical top gate bias V$_{TG}$= -18 V, we observe a clear noise in the modulation of the critical current which vanishes at V$_{TG}$= -30 V, well beyond the transition. The unusual features associated with these data such as non-zero critical current at the minima of the oscillations and its insensitivity to gate voltage are discussed elsewhere~\citep{Kurter2013}.

In summary, we report on gated Josephson devices with Bi$_2$Se$_3$ barriers in which the chemical potential can be tuned into different regimes. Our observations support the picture of a transition associated with the change in location of the topologically non-trivial low-energy ABSs that carry the majority of the supercurrent. Across this transition we find a qualitative change in the temperature dependence and phase coherent signatures of the supercurrent. When more than one magnetic flux quantum is threaded within the junction, we find that the gate-tuned transition is suppressed. As the transition is approached, low frequency fluctuations can cause the junction to switch prematurely to a low density behavior. The work expands the current understanding of topological insulators coupling to superconductors and nature of the proximity-induced supercurrent.

$\it{Acknowledgements.}$ CK, ADKF, and DJVH acknowledge funding by Microsoft Station-Q. For the device fabrication, we acknowledge use of the facilities of the Frederick Seitz Materials Research Laboratory at the University of Illinois at Urbana-Champaign.  PG acknowledges support from NSF DMR-1064319. YSH acknowledges support from National Science Foundation grant DMR-12-55607. We are thankful for helpful discussions with R.R.~Biswas, E.~Fradkin, T.~Hughes, V. Orlyanchik, and M. Stehno.

\bibliography{TriJunction}

\end{document}